\documentclass[aps,prl,reprint,superscriptaddress,floatfix]{revtex4-2}

\usepackage{graphicx}
\usepackage{bm}
\usepackage{dcolumn}
\usepackage[version=4]{mhchem}
\usepackage{siunitx}
\usepackage{braket}
\usepackage{booktabs}
\usepackage{amsmath}

\providecommand{\ket}[1]{|#1\rangle}
\providecommand{\bra}[1]{\langle#1|}

\begin{document}

\title{Absorption–emission quantum repeater using diamond quantum memories}

\newcommand{\YNU}{Department of Physics, Graduate School of Engineering Science, Yokohama National University, 79-5 Tokiwadai, Hodogaya-ku, Yokohama 240-8501, Japan}
\newcommand{\YNUF}{Faculty of Engineering, Yokohama National University, 79-5 Tokiwadai, Hodogaya-ku, Yokohama 240-8501, Japan}
\newcommand{\QIC}{Quantum Information Research Center, Institute of Advanced Sciences, Yokohama National University, 79-5, Tokiwadai, Hodogaya, Yokohama, 240-8501, Japan}
\newcommand{\AIST}{Advanced Power Electronics Research Center, National Institute of Advanced Industrial Science and Technology (AIST), 1-1-1 Umezono, Tsukuba, Ibaraki 305-8568, Japan}

\author{Taichi Fujiwara}
\affiliation{\YNU}

\author{Yuhei Sekiguchi}
\affiliation{\YNU}
\affiliation{\YNUF}
\affiliation{\QIC}

\author{Raustin Reyes}
\affiliation{\YNU}

\author{Toshiharu Makino}
\affiliation{\AIST}
\affiliation{\QIC}

\author{Hiromitsu Kato}
\affiliation{\AIST}
\affiliation{\QIC}

\author{Hideo Kosaka}
\email{kosaka-hideo-yp@ynu.ac.jp}
\affiliation{\YNU}
\affiliation{\YNUF}
\affiliation{\QIC}

\date{\today}

\begin{abstract}
Quantum repeaters are essential for overcoming the exponential photon loss that limits entanglement generation over long distances in quantum networks.
An absorption--emission--based quantum repeater exploits the fundamental light–matter interactions of a diamond nitrogen--vacancy (NV) center---photon absorption and photon emission---to transfer a quantum state from an absorbed photon to an emitted photon, offering a scalable architecture that operates without photon interference between remote nodes.
Here we demonstrate an absorption--emission--based quantum repeater node using a single NV center, realizing the complete single-node operation in which heralded photon-to-memory quantum state transfer, repeat-until-success (RUS) emission of a spin-entangled photon, and quantum teleportation of the memory state onto the emitted photon constitute the essential repeater operation.
By characterizing the complete repeater operation as a quantum channel from the absorbed photon to the emitted photon via quantum process tomography, we obtain a process fidelity of \qty{78}{\percent}.
This demonstration establishes the absorption--emission approach as a fundamental building block for scalable quantum repeater architectures and paves the way toward practical long-distance quantum networks.
\end{abstract}

\maketitle

\section{Introduction}
Long-distance connections between remote qubits enabled by quantum entanglement allow nonlocal operations and constitute a fundamental resource for quantum cryptography, quantum computation, and quantum sensing\cite{Kimble2008,Wehner2018,Awschalom2021}.
However, in realistic quantum networks, exponential photon loss with transmission distance severely limits the rate of entanglement generation over long distances.
Quantum repeaters have therefore been proposed as a promising approach to overcome this fundamental limitation\cite{Briegel1998,Duan2001,Azuma2023}.
Among various physical platforms, diamond color centers are particularly attractive for quantum repeater applications because they host multiple optically addressable quantum memories with long coherence times\cite{doherty2013nitrogen,bernien2013heralded,humphreys2018deterministic}.

An absorption--emission--based quantum repeater scheme transfers a quantum state from an absorbed photon to an emitted photon via the fundamental interactions of photon absorption and emission in a diamond nitrogen--vacancy (NV) center.
This photon-to-photon polarization state transfer process has recently been demonstrated experimentally\cite{Reyes2025_QT_absorb_emit}.
However, that demonstration relied on a emission attempt without a repeat-until-success (RUS) protocol, and therefore did not show the memory-enabled enhancement of the entanglement generation rate that defines a quantum repeater.
Furthermore, it did not use coherent zero-phonon-line emission and evaluated fidelities only for a limited set of basis states.
Related state-transfer functionalities have been pioneered with single atoms in optical cavities, including the reversible mapping of a photonic polarization qubit into and out of a single atom\cite{Specht2011}, its heralded storage\cite{Kalb2015}, and the transfer of a quantum state between remote atoms via the emission and absorption of a single photon\cite{Ritter2012}. Heralded absorption of a photonic qubit has also been realized with a charged quantum dot\cite{Delteil2017}.
Compared with interference-based schemes requiring precise optical synchronization\cite{Moehring2007,Hofmann2012,bernien2013heralded,humphreys2018deterministic,vanLeent2022,Hermans2023}, or scattering-based schemes relying on strong spin--photon coupling in nanophotonic cavities\cite{janitz2020cavity,Bhaskar2020,Knaut2024}, the absorption--emission approach relaxes these physical requirements.

Here, we demonstrate an absorption--emission--based quantum repeater node using a single NV center.
The node integrates heralded photon absorption with a memory-preserving repeat-until-success (RUS) protocol for coherent zero-phonon-line (ZPL) emission.
The RUS protocol increases the photon collection efficiency by approximately an order of magnitude relative to a single excitation attempt while preserving the stored quantum information.
We characterize the complete absorption-to-emission process as a quantum channel via quantum process tomography, obtaining a process fidelity that establishes genuine repeater-node functionality rather than a single attempt state transfer.
This work establishes the absorption--emission approach as a building block for scalable quantum repeater architectures.

\begin{figure*}[t]
    \centering
    \includegraphics[width=\linewidth]{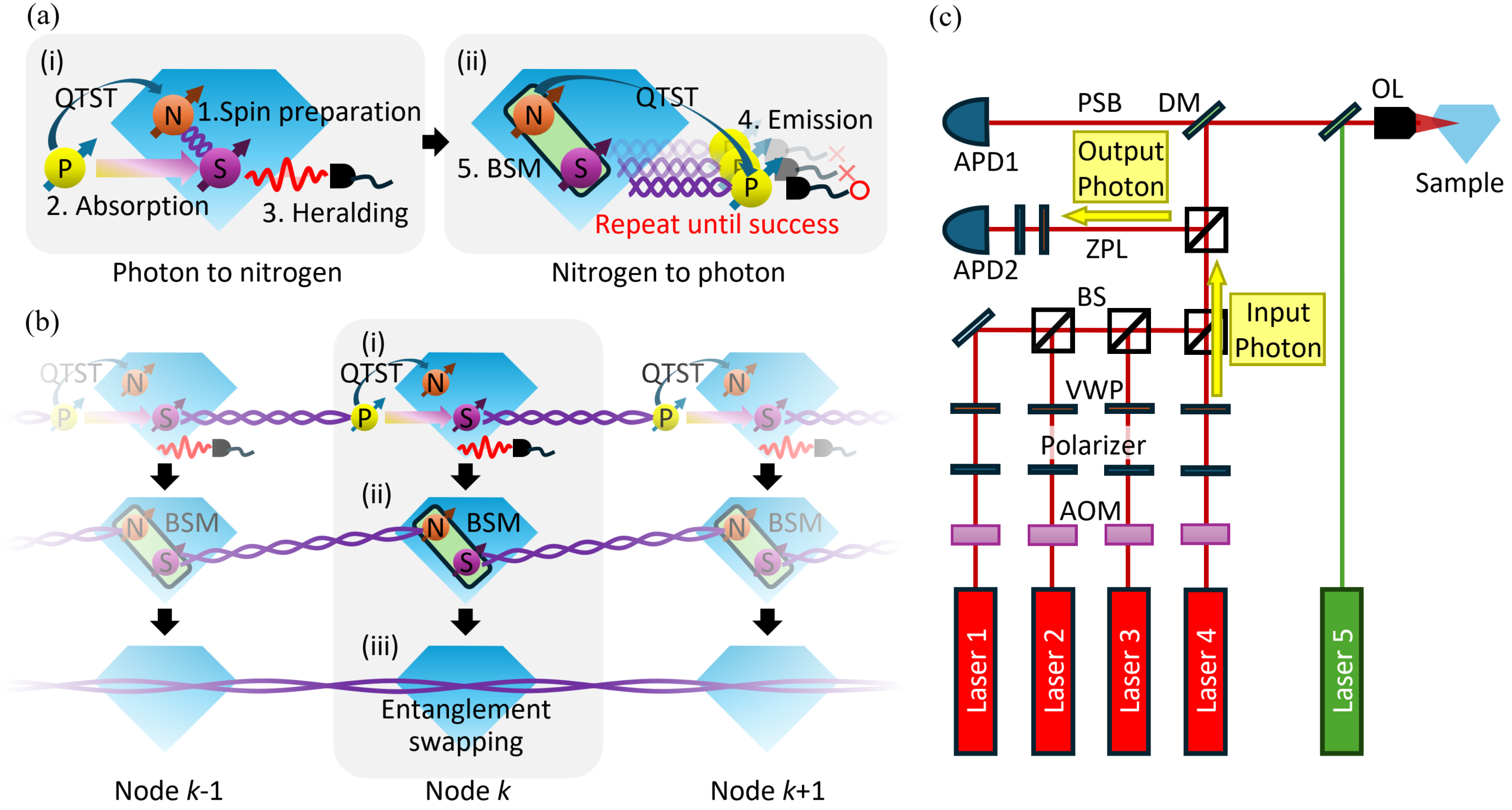}
    \caption{\textbf{Absorption--emission quantum repeater based on a single NV center.}
    (a) Operating principle of the repeater node. P, S, and N denote the photon polarization, the electron spin, and the nitrogen nuclear spin, respectively. (i) A polarization-encoded input photon is heraldedly absorbed by an NV center prepared in an electron--nuclear spin entangled state, which realizes quantum-teleportation-based state transfer (QTST) of the photonic state onto the nuclear-spin memory. (ii) Subsequent optical excitation generates a zero-phonon-line (ZPL) photon entangled with the electron spin. This emission process can be repeated until success. A local Bell-state measurement (BSM) between the electron and nuclear spins completes the end-to-end polarization state transfer to the emitted photon.
    (b) Conceptual extension to a multi-node quantum repeater architecture. (i) Sequential absorption--emission operations (QTST) transfer the state across nodes. (ii) Local BSMs at intermediate nodes allow the state-transfer scheme to be concatenated, resulting in (iii) entanglement swapping. The ZPL photon emitted from one node serves as the input photon for heralded absorption at the adjacent node.
    (c) Experimental implementation. A single NV center in diamond (sample) is operated at cryogenic temperature through an objective lens (OL). The optical setup includes multiple lasers, acousto-optic modulators (AOM), variable waveplates (VWP), polarizers, and beamsplitters (BS) to manipulate the spin state and input photon. Emitted photons are separated by a dichroic mirror (DM) into phonon-sideband (PSB) and ZPL components, which are detected by avalanche photodiodes (APD1 and APD2), respectively. The ZPL polarization is analyzed to characterize the output photon of the repeater channel.}
    \label{fig:concept}
\end{figure*}

\section{Principle of Operation}
The absorption--emission quantum repeater demonstrated in this work performs quantum-teleportation-based state transfer (QTST) between a photon polarization and a spin quantum memory via a single nitrogen--vacancy (NV) center in diamond (Fig.~\ref{fig:concept}a).
The transfer process exploits two ingredients, the spin--orbit coupling in the optically excited state and the hyperfine interaction between the electron and nitrogen nuclear spins in the ground state.

We label the relevant spin, orbital, and photonic states by the projection of angular momentum along the NV symmetry axis.
In this notation, $\ket{\pm 1}_\mathrm{S}$ and $\ket{\pm 1}_\mathrm{N}$ denote the electron-spin and the nitrogen nuclear-spin states with $S_z = \pm 1$ and $I_z = \pm 1$, respectively.
The optically excited orbital manifold is described by states $\ket{\pm 1}_\mathrm{L}$ with $L_z = \pm 1$, and photon polarization is expressed in the right- and left-handed circular polarization basis $\ket{\pm 1}_\mathrm{P}$\cite{Maze2011,Doherty2011}.

The protocol begins with the preparation of a maximally entangled state between the electron spin and the nitrogen nuclear spin in the orbital ground state,
\begin{equation}
    \ket{\Phi^+}_\mathrm{S,N} = \frac{1}{\sqrt{2}}\left( \ket{-1}_\mathrm{S}\otimes\ket{-1}_\mathrm{N} + \ket{+1}_\mathrm{S}\otimes\ket{+1}_\mathrm{N} \right),
\end{equation}
which is generated using microwave and radio-frequency control of the NV spin degrees of freedom\cite{Nagata2018}.
A polarization-encoded input photon in an arbitrary state $\ket{\psi}_{\mathrm{P_{in}}} = \alpha\ket{+1}_{\mathrm{P_{in}}} + \beta\ket{-1}_{\mathrm{P_{in}}}$ is subsequently absorbed on an optical transition resonant with the excited orbital state
\begin{equation}
    \ket{A_2}_\mathrm{L,S} = \frac{1}{\sqrt{2}}\left( \ket{-1}_\mathrm{L}\otimes\ket{+1}_\mathrm{S} + \ket{+1}_\mathrm{L}\otimes\ket{-1}_\mathrm{S} \right).
\end{equation}
Owing to angular-momentum conservation between the electronic orbital and the photon polarization, the observation of a photon absorption event constitutes an effective Bell-state projection onto
\begin{equation}
    \ket{\Psi^+}_{\mathrm{P_{in}},\mathrm{S}} = \frac{1}{\sqrt{2}}\left( \ket{-1}_{\mathrm{P_{in}}}\otimes\ket{+1}_\mathrm{S} + \ket{+1}_{\mathrm{P_{in}}}\otimes\ket{-1}_\mathrm{S} \right)
\end{equation}
in the joint electron-spin--photon system\cite{Kosaka2015}.
Conditioned on this heralded absorption event, the photonic polarization state is transferred onto the nitrogen nuclear spin via quantum teleportation\cite{Yang2016,tsurumoto2019quantum, Ito2025}.
The resulting nuclear-spin state is given by $\sigma_x\ket{\psi}_\mathrm{N}$.

In the second state transfer step, the quantum state stored in the nitrogen nuclear-spin memory is transferred to the polarization state of an emitted photon.
The NV center is optically excited to $\ket{A_2}_\mathrm{L,S}$ state, leading to the emission of a photon on the zero-phonon line (ZPL), where the photon polarization is entangled with the electron spin, $\ket{\Psi^+}_{\mathrm{P_{out}},\mathrm{S}}$\cite{togan2010quantum,Sekiguchi2021}.
Since the entangled emission itself does not involve the nuclear-spin memory, a repeat-until-success (RUS) emission can be implemented.
Conditioned on a successful emission, the state of the entire system is described as
\begin{align}
\ket{\Psi^+}_{\mathrm{P_{out}},\mathrm{S}} \otimes         \sigma_x\ket{\psi}_\mathrm{N}
    =\ &\frac{1}{2}\Big(\ket{\psi}_\mathrm{P_{out}}\otimes\ket{\Phi^+}_\mathrm{S,N} \nonumber\\
    &- \sigma_z\ket{\psi}_\mathrm{P_{out}}\otimes\ket{\Phi^-}_\mathrm{S,N} \nonumber\\
    &+ \sigma_x\ket{\psi}_\mathrm{P_{out}}\otimes\ket{\Psi^+}_\mathrm{S,N} \nonumber\\
    &+ \sigma_x\sigma_z\ket{\psi}_\mathrm{P_{out}}\otimes\ket{\Psi^-}_\mathrm{S,N}\Big).
\end{align}
A Bell-state measurement (BSM) between the electron spin and the nitrogen nuclear spin is then performed.
Conditioned on successful projection onto the Bell state $\ket{\Phi^+}_\mathrm{S,N}$, the nuclear-spin state is transferred onto the polarization of the emitted photon, completing the absorption--emission quantum channel.
While this Bell measurement can in principle be implemented deterministically\cite{Reyes2022,Kamimaki2023}, the present experiment employs a probabilistic projection to ensure high-fidelity state transfer.

This single-node operation serves as the elementary building block of a scalable multi-node repeater chain, in which concatenated state transfer and local Bell-state measurements at intermediate nodes realize entanglement swapping across the network (Fig.~\ref{fig:concept}b).

\begin{figure*}[t]
    \centering
    \includegraphics[width=\linewidth]{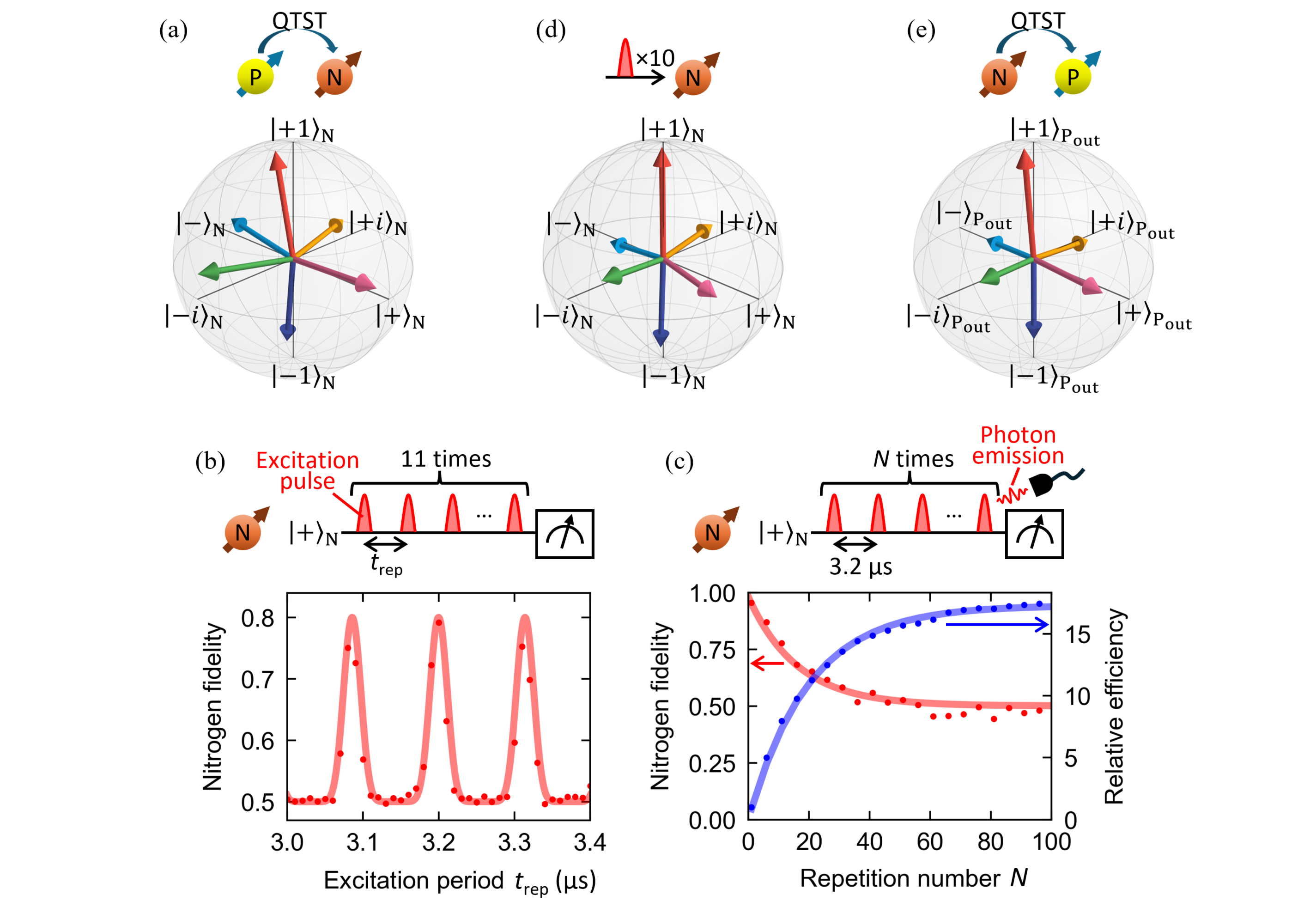}
    \caption{\textbf{Memory-preserving repeat-until-success operation and elementary state-transfer processes.}
    (a) Quantum state transfer from an absorbed photon to the nitrogen nuclear spin. For six input photonic polarization states, the transferred nuclear-spin states are reconstructed by quantum state tomography, demonstrating high-fidelity photon-to-spin state transfer.
    (b) Synchronization of the excitation repetition period. The nitrogen nuclear spin is initialized in a superposition state and subjected to eleven repeated optical excitations, and the nuclear-spin fidelity after the eleventh excitation is measured as a function of the repetition period $t_{\mathrm{rep}}$, revealing an optimal value that preserves nuclear-spin coherence.
    (c) Trade-off between nuclear-spin state preservation and photon collection efficiency as a function of the number of excitation attempts $N$. The nuclear-spin fidelity (red) is evaluated after the $N$-th attempt and decreases with increasing $N$ due to residual dephasing, while the relative photon collection efficiency (blue), accumulated up to the $N$-th attempt, increases.
    In (b) and (c), the reported nuclear-spin fidelities are corrected for independently characterized state-preparation-and-measurement (SPAM) errors.
    (d) Preservation of arbitrary nuclear-spin states under repeat-until-success (RUS) emission. Six nuclear-spin basis states are subjected to ten excitation attempts, and are reconstructed by quantum state tomography, yielding an average fidelity of \qty{87}{\percent}.
    (e) Quantum state transfer from the nuclear spin to the polarization of an emitted photon. Following RUS-based ZPL emission and a local Bell-state measurement, the polarization states of the emitted photons are reconstructed, confirming spin-to-photon state transfer.
    The points represent experimentally measured data, and the solid lines are a fit to the data (see Supplementary Note~\ref{supp:coherence}).}
    \label{fig:transfer}
\end{figure*}

\section{Experiments and Results}

\subsection{Photon-to-nuclear-spin state transfer}
All experiments are performed on a single NV center in diamond using the optical setup shown in Fig.~\ref{fig:concept}c (see Methods). We first verify the quantum state transfer from an absorbed photon to the nitrogen nuclear spin.
The NV center is initialized in an entangled state between the electron spin and the nitrogen nuclear spin, after which a weak optical pulse, prepared in one of the six polarization basis states $\{\ket{\pm 1}_{\mathrm{P_{in}}}, \ket{\pm}_{\mathrm{P_{in}}}, \ket{\pm i}_{\mathrm{P_{in}}}\}$, is resonantly absorbed, where $\ket{\pm} = (\ket{+1} \pm \ket{-1})/\sqrt{2}$ and $\ket{\pm i} = (\ket{+1} \pm i\ket{-1})/\sqrt{2}$ for each degree of freedom.
Successful absorption is heralded by detection of phonon-sideband (PSB) emission.

Conditioned on the absorption herald, we measure the nitrogen nuclear spin in the six orthogonal spin basis states, $\{\ket{\pm 1}_\mathrm{N}, \ket{\pm}_\mathrm{N}, \ket{\pm i}_\mathrm{N}\}$.
For each input photonic polarization state, the transferred nuclear-spin state is reconstructed using quantum state tomography.
From the reconstructed density matrices, we obtain an average state fidelity of \qty{93}{\percent}, demonstrating high-fidelity photon-to-nuclear-spin quantum state transfer (Fig.~\ref{fig:transfer}a).

\subsection{Memory-preserving emission process}
NV nuclear spins are only weakly perturbed by optical cycling of the electron spin, which enables repetitive and single-shot spin readout\cite{Jiang2009,Neumann2010,Robledo2011}.
However, maintaining coherence across repeated excitations is much more challenging. Moreover, if the electron and nuclear spins are entangled when a photon entangled with the electron spin is lost, the nuclear-spin state is destroyed as well.
In the orbital ground state, the hyperfine interaction periodically entangles and disentangles the electron and nuclear spins. Exciting at the moments of disentanglement protects the nuclear-spin state against photon loss. We therefore synchronize the excitation interval with the hyperfine period (see Supplementary Note~\ref{supp:coherence}). 
The dephasing of nuclear-spin memories under repeated optical operations has also been studied in the context of remote entanglement generation\cite{Reiserer2016,Kalb2018,Bradley2022}.

Optical excitation for entangled emission is performed using \qty{3}{\nano\second} $\pi$-pulses resonant with the $A_2$ optical transition, with a pulse power of \qty{1.2}{\micro\watt}.
The excitation sequence is repeated periodically until a zero-phonon-line (ZPL) photon is successfully detected, which heralds completion of the emission step.
We first calibrate the optimal excitation repetition period $t_{\mathrm{rep}}$ by initializing the nuclear spin in the superposition state $\ket{+}_\mathrm{N}$ and measuring the population of this state conditioned on the ZPL photon detection after eleven excitation attempts, as a function of the repetition period $t_{\mathrm{rep}}$.
From this measurement, we identify an optimal repetition period of $t_{\mathrm{rep}} = \qty{3.2}{\micro\second}$, which maximizes nuclear-spin coherence during repeated excitation (Fig.~\ref{fig:transfer}b). 

Using this calibrated repetition period, we investigate the trade-off between nuclear-spin state preservation and photon collection efficiency as a function of the number of excitation attempts $N$.
For each $N$, we condition on detection of a ZPL photon after the $N$th attempt and measure the population of the $\ket{+}_\mathrm{N}$ state.
The observed decrease in fidelity with increasing $N$ reflects residual dephasing arising from stochastic timing jitter in the excited-state relaxation process (Fig.~\ref{fig:transfer}c) (see Supplementary Note~\ref{supp:coherence}).
Based on this trade-off, we limit the maximum number of excitation attempts to ten in the subsequent experiments, ensuring relatively high nuclear-spin fidelity while benefiting from the increased emission success probability.

We then perform full state tomography on the nuclear-spin states $\{\ket{\pm 1}_\mathrm{N}, \ket{\pm}_\mathrm{N}, \ket{\pm i}_\mathrm{N}\}$, conditioned on ZPL detection at the tenth attempt.
The average fidelity is \qty{87}{\percent}, showing that the nuclear-spin memory retains an arbitrary state after the RUS emission (Fig.~\ref{fig:transfer}d).
The dominant error is dephasing of the superposition states.

\subsection{Nuclear spin-to-photon polarization state transfer}
We next characterize the QTST from the nitrogen nuclear spin to the polarization of an emitted photon.
The nuclear spin is initialized in one of six basis states $\{\ket{\pm 1}_\mathrm{N}, \ket{\pm}_\mathrm{N}, \ket{\pm i}_\mathrm{N}\}$, after which optical excitation resonant with the transition is applied to generate a polarization-entangled photon in the zero-phonon line (ZPL).
The emission operation is implemented using the RUS protocol described above.
Following successful detection of a ZPL photon, a Bell measurement between the electron spin and the nuclear spin is performed.
While a deterministic Bell measurement between the electron and nuclear spins is in principle feasible\cite{Reyes2022,Kamimaki2023}, we employ a probabilistic projection onto the $\ket{\Phi^+}_\mathrm{S,N}$ state to ensure high-fidelity state transfer.

The emitted photon is analyzed in the six basis states $\{\ket{\pm 1}_\mathrm{P_{out}}, \ket{\pm}_\mathrm{P_{out}}, \ket{\pm i}_\mathrm{P_{out}}\}$.
Using the measured joint statistics conditioned on successful Bell-measurement outcomes, we reconstruct the polarization density matrices of the emitted photons via quantum state tomography (see Supplementary Note~\ref{supp:tomography}).
We obtain an average state fidelity of \qty{83}{\percent} across the six input nuclear-spin states (Fig.~\ref{fig:transfer}e).
This value is mainly limited by imperfect preservation of the nuclear-spin state during repeated optical excitation.

\begin{figure*}[t]
    \centering
    \includegraphics[width=\linewidth]{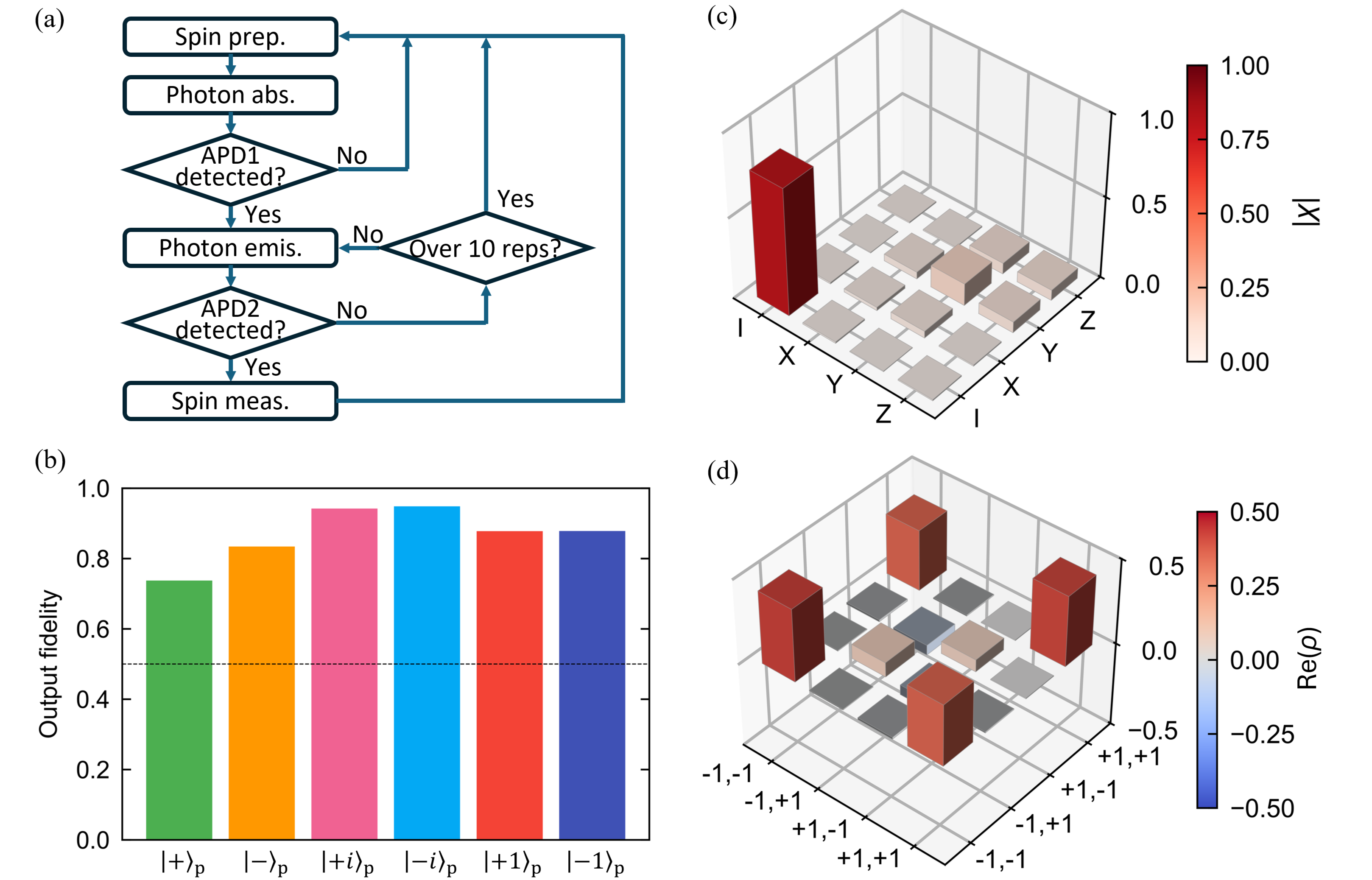}
    \caption{\textbf{Demonstration of absorption--emission quantum repeater--node operation.}
    (a) Experimental sequence for the complete repeater-node operation. Following probabilistic initialization of the NV-center charge state, the spin states are prepared (Spin prep.) and a resonant photon is absorbed (Photon abs.). Successful absorption is heralded by detection of a relaxation photon, followed by repeat-until-success generation of an entangled photon and a local Bell-state ($\ket{\Phi^+}_\mathrm{S,N}$) measurement between the electron and nuclear spins (Spin meas.).
    (b) Output-state fidelity of the emitted photon for six input polarization states.
    (c) Absolute values of the process matrix elements $|\chi_{ij}|$, reconstructed from the experimental data. The process fidelity is \qty{78}{\percent} with respect to the identity operation, demonstrating photon-to-photon quantum state transfer mediated by the NV-center repeater node.
    (d) Real part of the corresponding Choi matrix representation of the process, $\rho_\text{choi}$ in the input-output polarization basis.}
    \label{fig:repeater}
\end{figure*}

\subsection{Demonstration of repeater-node operation}
Finally, we demonstrate the complete operation of the absorption--emission quantum repeater node by combining photon absorption, photon emission with RUS, and Bell measurement within a single experimental sequence (Fig.~\ref{fig:repeater}a).
The NV center is initialized in an entangled state $\ket{\Phi^+}_\mathrm{S,N}$.
A weak optical pulse, prepared in one of the six polarization basis states $\{\ket{\pm 1}_{\mathrm{P_{in}}}, \ket{\pm}_{\mathrm{P_{in}}}, \ket{\pm i}_{\mathrm{P_{in}}}\}$, is then resonantly absorbed, with successful absorption heralded by PSB detection.
Following the absorption event, optical excitation is applied to generate a ZPL photon using the RUS protocol.
Upon successful detection of the emitted photon, a Bell measurement between the electron spin and the nitrogen nuclear spin is performed.
The emitted photon is analyzed in one of the six polarization bases $\{\ket{\pm 1}_\mathrm{P_{out}}, \ket{\pm}_\mathrm{P_{out}}, \ket{\pm i}_\mathrm{P_{out}}\}$, enabling reconstruction of the photonic output state conditioned on successful Bell-measurement outcomes (Fig.~\ref{fig:repeater}b).
The implementation of the Bell measurement and the analysis method are the same as in Ref\cite{Reyes2025_QT_absorb_emit}.

Using the reconstructed output states for all six input polarizations, we characterize the absorption--emission--based repeater operation as a quantum channel by performing quantum process tomography.
The resulting process matrix is shown in Fig.~\ref{fig:repeater}c,d, from which we extract a process fidelity of \qty{78}{\percent} with respect to the identity operation.
Known unitary rotations arising from the Bell-measurement outcomes are compensated in post-processing, reflecting the fact that these operations can be corrected deterministically using classical feedforward in a practical repeater implementation.

\section{Discussion}
In this work, we have demonstrated an absorption--emission--based quantum repeater node using a single diamond NV center.
The demonstrated node integrates heralded photon absorption, memory-preserving repeat-until-success (RUS) emission, and end-to-end characterization of the polarization-encoded quantum channel via quantum process tomography.
By explicitly verifying the absorption--emission process as a non-classical quantum channel with a process fidelity of \qty{78}{\percent}, this work establishes the experimental feasibility of such architectures as a fundamental primitive for future quantum networks.

A key feature of the absorption--emission approach is that it relies neither on photon interference between remote nodes nor on strong spin--photon coupling in optical cavities.
In contrast to interference-based repeater schemes, which require precise temporal, spectral, and spatial mode matching of photons, the present protocol operates through a local quantum interface between a flying photonic qubit and a stationary spin quantum memory.
This locality enables an interference-free and cavity-free repeater operation, relaxing critical implementation requirements while preserving the essential functionality of a quantum repeater node.

The repeat-until-success emission protocol demonstrated here plays a central role in enabling repeater operation.
Rather than serving merely as a technical method to improve emission probability, the RUS protocol constitutes a memory-enabled conditional photonic interface, which is a key element for quantum repeater architectures that rely on conditional emission strategies to mitigate photon-loss effects.
The ability of the nuclear-spin quantum memory to preserve the stored quantum state during repeated optical excitation attempts is therefore a defining requirement for repeater functionality.
By experimentally verifying memory preservation during RUS emission, this work demonstrates that the absorption--emission process can operate as a genuine repeater node rather than a standalone state-transfer device.

The performance of the demonstrated repeater node is currently limited primarily by imperfect preservation of the nuclear-spin quantum memory during repeated excitation, which is mainly attributed to stochastic timing jitter in the relaxation dynamics of the optically excited state and the associated hyperfine interaction.
Crystal strain in the optically excited state, in contrast, is not a significant contribution under the strain conditions of the present NV center (Supplementary Note~\ref{supp:strain}).
In the present experiment, the number of RUS attempts was limited to ten in order to maintain high-fidelity state transfer. Nevertheless, this repetition number can already be sufficient to provide a rate advantage in parameter regimes characterized by high absorption--emission efficiency and low direct transmission efficiency (see Supplementary Note~\ref{supp:efficiency}).

Beyond the present implementation, the demonstrated absorption--emission protocol can be naturally extended through improvements at both the device and protocol levels, without modifying its underlying operating principle.
At the protocol level, further improvements may be achieved through excitation sequences designed to suppress nuclear-spin dephasing during repeated optical excitation, thereby relaxing the current limitation on the number of repeat-until-success attempts\cite{Reiserer2016, Bradley2022}. 
At the device level, direct photonic integration between diamond structures and optical fibers offers a clear route toward improved collection and coupling efficiency.
In addition, further performance gains may be obtained by engineering the local photonic environment, for example through moderate Purcell enhancement of zero-phonon-line emission\cite{janitz2020cavity,Ruf2021,Yurgens2024,Ding2024}.

Moreover, this repetitive emission process could also serve as a route toward the sequential generation of photonic cluster states from a single emitter\cite{Lindner2009,Buterakos2017,Russo2018}, suggesting its potential for graph-state resources for all-photonic quantum repeaters and fault-tolerant photonic quantum communication.

More generally, the absorption--emission repeater concept applies to quantum systems that combine (i) a degenerate ground-state doublet, (ii) circularly polarized optical selection rules coupling both qubit states to a common spin--orbit entangled excited state, and (iii) a long-lived auxiliary quantum memory.
These requirements are naturally fulfilled by neutral atoms with degenerate Zeeman levels\cite{Ritter2012,Specht2011,Kalb2015} and are partially met by charged quantum dots through trion transitions\cite{Delteil2017}; extending the scheme to other solid-state emitters would require engineering of the corresponding optical selection rules.
The zero-field NV center thus constitutes a solid-state platform in which all three requirements coexist without an optical cavity, providing a concrete framework for constructing scalable quantum repeater networks based on photonic quantum channels mediated by local, memory-assisted quantum interfaces.

\section{Methods}
The NV center used in this experiment is a naturally occurring defect in an electronic-grade diamond substrate (Element Six) with $\langle 111\rangle$ crystal orientation. A solid immersion lens (SIL) with a diameter of \qty{2}{\micro\meter} is fabricated on the surface of the substrate\cite{Hadden2010,Siyushev2010}, and an anti-reflection coating of \qty{118}{\nano\meter}-thick \ce{SiO2} is deposited on top of the SIL. The sample is cooled to \qty{5.5}{K} and integrated with a cross-wire microwave antenna for spin control. Residual static magnetic fields, including the geomagnetic field, are compensated using a three-axis coil system, maintaining the NV center at zero magnetic field throughout the experiment.

The optical setup used for spin control, heralded absorption, and photon detection is shown in Fig.~\ref{fig:concept}c.
The charge state of the NV center is probabilistically initialized using a \qty{515}{\nano\meter} laser (Laser 5). The electron spin is initialized into the $\ket{0}_\mathrm{S}$ state using Laser 1, which is resonant with the $\ket{E_{1,2}}_\mathrm{L,S}$ optical transitions. State-selective excitation of the electron spin is performed using Laser 2, resonant with the $\ket{E_x}_\mathrm{L,S}$ transition, which selectively excites the $\ket{0}_\mathrm{S}$ state to the $\ket{E_x}_\mathrm{L,S}$ excited state. The electron-spin state is then read out by detecting phonon-sideband (PSB) emission with a single-photon avalanche photodiode (APD1)\cite{Robledo2011}. Initialization of the nitrogen nuclear spin, and generation of the electron--nuclear spin entangled state $\ket{\Phi^+}_\mathrm{S,N}$, are performed using microwave and radio-frequency control following the same protocol as in Ref.~\cite{Nagata2018}.

Optical excitation for entangled ZPL emission is performed using Laser 3, which is resonant with the $\ket{A_2}_\mathrm{L,S}$ transition and pulsed into \qty{3}{\nano\second} $\pi$-pulses with a power of \qty{1.2}{\micro\watt} by an acousto-optic modulator (AOM).
The polarization-encoded input optical pulses used for heralded absorption are derived from Laser 4.
The optical power of Laser 4 is attenuated using a variable attenuator to produce weak coherent pulses, and its polarization is set to an arbitrary state using two variable waveplates (VWP).
The resulting continuous-wave light is converted into a pulsed input photon using an AOM. 
The input pulse has a duration of \qty{12}{\nano\second} and a mean photon number of ~400 at the sample.
Since an absorption efficiency is $\sim10^{-3}$, the absorption events are dominated by single-photon absorption.

Emitted photons are separated into ZPL and PSB components using a dichroic mirror (DM). The ZPL component is selected using a narrowband bandpass filter centered at \qty{637}{\nano\meter} with a full width at half maximum (FWHM) of \qty{5}{\nano\meter}. The PSB component is isolated using a combination of long-pass and short-pass filters, transmitting the \qtyrange{650}{780}{\nano\meter} wavelength range.

The repeat-until-success (RUS) protocol is implemented using a home-built field-programmable gate array (FPGA) system, which serves as the master controller. The FPGA performs real-time conditional branching based on photon-detection signals, determining whether to repeat the excitation sequence or proceed to the next step. An arbitrary waveform generator (AWG) generates the spin-control waveforms and switches between control sequences according to the branching decisions issued by the FPGA.

The Bell-state measurement between the electron spin and the nitrogen nuclear spin is implemented following the approach of Ref.~\cite{Reyes2025_QT_absorb_emit}: microwave and radio-frequency pulses transfer the population of the target Bell state $\ket{\Phi^+}_\mathrm{S,N}$ to the joint state $\ket{0,0}_\mathrm{S,N}$, and successful projection is confirmed by reading out the $\ket{0}_\mathrm{N}$ state of the nuclear spin.

\section{Acknowledgments}
We acknowledge support from the Japan Science and Technology Agency (JST) Moonshot R\&D (JPMJMS2062), JST CREST (JPMJCR1773), and JST ASPIRE (JPMJAP24C1).
We also acknowledge support from the Ministry of Internal Affairs and Communications R\&D of ICT Priority Technology Project (JPMI00316) and JSPS KAKENHI (20H05661, 20K20441, 25H00830).

\section*{Data Availability}
The data that support the findings of this study are available from the corresponding author upon reasonable request.

\bibliography{references}

@article{Kimble2008,
  title   = {The quantum internet},
  author  = {Kimble, H. J.},
  journal = {Nature},
  volume  = {453},
  number  = {7198},
  pages   = {1023--1030},
  year    = {2008},
  doi     = {10.1038/nature07127}
}

@article{Wehner2018,
  title = {Quantum internet: A vision for the road ahead},
  author = {Wehner, Stephanie and Elkouss, David and Hanson, Ronald},
  journal = {Science},
  volume = {362},
  number = {6412},
  pages = {eaam9288},
  year = {2018},
  publisher = {American Association for the Advancement of Science},
  doi = {10.1126/science.aam9288}
}

@article{Awschalom2021,
  title = {Development of Quantum Interconnects (QuICs) for Next-Generation Information Technologies},
  author = {Awschalom, D. D. and Hanson, R. and Wrachtrup, J. and Zhou, B. B.},
  journal = {PRX Quantum},
  volume = {2},
  number = {1},
  pages = {017002},
  year = {2021},
  publisher = {American Physical Society},
  doi = {10.1103/PRXQuantum.2.017002}
}

@article{Azuma2023,
  title = {Quantum repeaters: From quantum networks to the quantum internet},
  author = {K. Azuma and S. E. Economou and D. Elkouss and P. Hilaire and L. Jiang and H.-K. Lo and I. Tzitrin},
  journal = {Reviews of Modern Physics},
  volume = {95},
  number = {4},
  pages = {045006},
  year = {2023},
  publisher = {American Physical Society},
  doi = {10.1103/RevModPhys.95.045006}
}

@article{Duan2001,
  author  = {Duan, L.-M. and Lukin, M. D. and Cirac, J. I. and Zoller, P.},
  title   = {Long-distance quantum communication with atomic ensembles and linear optics},
  journal = {Nature},
  volume  = {414},
  pages   = {413--418},
  year    = {2001},
  doi     = {10.1038/35106500}
}

@article{doherty2013nitrogen,
  title={The nitrogen-vacancy colour centre in diamond},
  author={Doherty, Marcus W and Manson, Neil B and Delaney, Paul and Jelezko, Fedor and Wrachtrup, J{\"o}rg and Hollenberg, Lloyd C. L.},
  journal = {Physics Reports},
  volume  = {528},
  number  = {1},
  pages   = {1--45},
  year    = {2013},
  doi      = {10.1016/j.physrep.2013.02.001},
  publisher={Elsevier}
}

@article{humphreys2018deterministic,
  title={Deterministic delivery of remote entanglement on a quantum network},
  author={Humphreys, P. C. and Kalb, N and Morits, J. P. J. and Schouten, R. N. and Vermeulen, R. F. L. and Twitchen, D. J. and Markham, M and Hanson, R},
  journal={Nature},
  volume={558},
  number={7709},
  pages={268--273},
  year={2018},
  doi={10.1038/s41586-018-0200-5},
  publisher={Nature Publishing Group}
}

@article{bernien2013heralded,
  title={Heralded entanglement between solid-state qubits separated by three metres},
  author={Bernien, H and Hensen, B and Pfaff, W and Koolstra, G and Blok, M. S. and Robledo, L and Taminiau, T. H. and Markham, M and Twitchen, D. J. and Childress, L and Hanson, R},
  doi={10.1038/nature12016},
  journal={Nature},
  volume={497},
  number={7447},
  pages={86--90},
  year={2013},
  publisher={Nature Publishing Group}
}

@article{Hermans2023,
  author  = {Hermans, S. L. N. and Pompili, M. and Dos Santos Martins, L. and Montblanch, A. R.-P. and Beukers, H. K. C. and Baier, S. and Borregaard, J. and Hanson, R.},
  title   = {Entangling remote qubits using the single-photon protocol: an in-depth theoretical and experimental study},
  journal = {New Journal of Physics},
  volume  = {25},
  number  = {1},
  pages   = {013011},
  year    = {2023},
  doi     = {10.1088/1367-2630/acb004}
}

@article{Kosaka2015,
  author  = {Kosaka, Hideo and Niikura, Naeko},
  title   = {Entangled absorption of a single photon with a single spin in diamond},
  journal = {Physical Review Letters},
  volume  = {114},
  number  = {5},
  pages   = {053603},
  year    = {2015},
  doi     = {10.1103/PhysRevLett.114.053603}
}

@article{Yang2016,
  author  = {S. Yang and Y. Wang and D. D. B. Rao and T. H. Tran and S. A. Momenzadeh and M. Markham and D. J. Twitchen and P. Wang and W. Yang and R. Stöhr and P. Neumann and H. Kosaka and  J. Wrachtrup},
  title   = {High-fidelity transfer and storage of photon states in a single nuclear spin},
  journal = {Nature Photonics},
  volume  = {10},
  pages   = {507--511},
  year    = {2016},
  doi     = {10.1038/nphoton.2016.103}
}

@article{tsurumoto2019quantum,
  title={Quantum teleportation-based state transfer of photon polarization into a carbon spin in diamond},
  author={Tsurumoto, Kazuya and Kuroiwa, Ryota and Kano, Hiroki and Sekiguchi, Yuhei and Kosaka, Hideo},
  journal={Communications Physics},
  volume={2},
  number={1},
  pages={74},
  year={2019},
  doi={10.1038/s42005-019-0158-0},
  publisher={Nature Publishing Group}
}

@article{Ito2025,
  title        = {Robust transfer of a quantum state from an absorbed photon into a diamond spin},
  author       = {Ito, Daisuke and Sekiguchi, Yuhei and Reyes, Raustin and Fujiwara, Taichi and Makino, Toshiharu and Kato, Hiromitsu and Kosaka, Hideo},
  journal      = {Opt. Lett.},
  volume       = {50},
  number       = {16},
  pages        = {5073--5076},
  year         = {2025},
  doi          = {10.1364/OL.567933}
}

@article{togan2010quantum,
  title={Quantum entanglement between an optical photon and a solid-state spin qubit},
  author={Togan, Emre and Chu, Yiwen and Trifonov, Andrei S and Jiang, Liang and Maze, Jared R and Childress, Lilian and Dutt, M V Gurudev and Sørensen, Anders S and Hemmer, Philip R and Zibrov, Alexander S and Lukin, M. D.},
  journal={Nature},
  volume={466},
  number={7307},
  pages={730--734},
  year={2010},
  doi={10.1038/nature09256},
  publisher={Nature Publishing Group}
}

@article{Sekiguchi2021,
  title        = {Geometric entanglement of a photon and spin qubits in diamond},
  author       = {Sekiguchi, Yuhei and Yasui, Yuki and Tsurumoto, Kazuya and Koga, Yuta and Reyes, Raustin and Kosaka, Hideo},
  journal      = {Communications Physics},
  volume       = {4},
  pages        = {264},
  year         = {2021},
  doi          = {10.1038/s42005-021-00767-1}
}

@article{Kamimaki2023,
  author  = {Kamimaki, Akira and Wakamatsu, Keidai and Mikata, Kosuke and Sekiguchi, Yuhei and Kosaka, Hideo},
  title   = {Deterministic Bell state measurement with a single quantum memory},
  journal = {npj Quantum Information},
  volume  = {9},
  pages   = {101},
  year    = {2023},
  doi     = {10.1038/s41534-023-00771-z}
}

@article{Knaut2024,
  title        = {Entanglement of Nanophotonic Quantum Memory Nodes in a Telecom Network},
  author       = {Knaut, C. M. and Suleymanzade, A. and Wei, Y.-C. and Assumpcao, D. R. and Stas, P.-J. and Huan, Y. Q. and Machielse, B. and Knall, E. N. and Sutula, M. and Baranes, G. and Sinclair, N. and De-Eknamkul, C. and Levonian, D. S. and Bhaskar, M. K. and Park, Hongkun and Lončar, Marko and Lukin, M. D.},
  journal      = {Nature},
  volume       = {629},
  number       = {8012},
  pages        = {573--578},
  year         = {2024},
  doi          = {10.1038/s41586-024-07252-z},
  publisher    = {Nature Publishing Group}
}

@article{Reyes2025_QT_absorb_emit,
  title   = {Quantum teleportation of a photon via absorption and emission for quantum repeater nodes},
  author  = {Reyes, Raustin and Sekiguchi, Yuhei and Ito, Daisuke and Fujiwara, Taichi and Watanabe, Kansei and Makino, Toshiharu and Kato, Hiromitsu and Kosaka, Hideo},
  journal = {npj Quantum Information},
  volume  = {12},
  number  = {1},
  pages   = {25},
  year    = {2026},
  doi     = {10.1038/s41534-025-01169-9}
}

@article{janitz2020cavity,
  title={Cavity quantum electrodynamics with color centers in diamond},
  author={Janitz, E and Bhaskar, M. K. and Childress, L},
  journal={Optica},
  volume={7},
  number={10},
  pages={1232--1252},
  year={2020},
  doi={10.1364/OPTICA.398628},
  publisher={Optica Publishing Group}
}

@article{Yurgens2024,
  author  = {Yurgens, Viktoria and Fontana, Yannik and Corazza, Andrea and Shields, Brendan J. and Maletinsky, Patrick and Warburton, Richard J.},
  title   = {Cavity-assisted resonance fluorescence from a nitrogen-vacancy center in diamond},
  journal = {npj Quantum Information},
  volume  = {10},
  number  = {1},
  pages   = {112},
  year    = {2024},
  doi     = {10.1038/s41534-024-00915-9}
}

@article{Ding2024,
  author  = {Ding, Sophie W. and Haas, Michael and Guo, Xinghan and Kuruma, Kazuhiro and Jin, Chang and Li, Zixi and Awschalom, David D. and Delegan, Nazar and Heremans, F. Joseph and High, Alexander A. and Lončar, Marko},
  title   = {High-Q cavity interface for color centers in thin film diamond},
  journal = {Nature Communications},
  volume  = {15},
  number  = {1},
  pages   = {6358},
  year    = {2024},
  doi     = {10.1038/s41467-024-50667-5}
}

@article{Nagata2018,
  author    = {Nagata, Kodai and Kuramitani, Kouyou and Sekiguchi, Yuhei and Kosaka, Hideo},
  title     = {Universal holonomic quantum gates over geometric spin qubits with polarised microwaves},
  journal   = {Nature Communications},
  year      = {2018},
  volume    = {9},
  number    = {1},
  pages     = {3227},
  doi       = {10.1038/s41467-018-05664-w},
  url       = {https://doi.org/10.1038/s41467-018-05664-w}
}

@article{Briegel1998,
  author  = {Briegel, H.-J. and D{\"u}r, W. and Cirac, J. I. and Zoller, P.},
  title   = {Quantum Repeaters: The Role of Imperfect Local Operations in Quantum Communication},
  journal = {Phys. Rev. Lett.},
  volume  = {81},
  pages   = {5932},
  year    = {1998},
  doi     = {10.1103/PhysRevLett.81.5932}
}

@article{Moehring2007,
  author  = {Moehring, D. L. and Maunz, P. and Olmschenk, S. and Younge, K. C. and Matsukevich, D. N. and Duan, L.-M. and Monroe, C.},
  title   = {Entanglement of Single-Atom Quantum Bits at a Distance},
  journal = {Nature},
  volume  = {449},
  pages   = {68},
  year    = {2007},
  doi     = {10.1038/nature06118}
}

@article{Hofmann2012,
  author  = {Hofmann, J. and Krug, M. and Ortegel, N. and G{\'e}rard, L. and Weber, M. and Rosenfeld, W. and Weinfurter, H.},
  title   = {Heralded Entanglement Between Widely Separated Atoms},
  journal = {Science},
  volume  = {337},
  pages   = {72},
  year    = {2012},
  doi     = {10.1126/science.1221856}
}

@article{vanLeent2022,
  author  = {van Leent, T. and Bock, M. and Fertig, F. and Garthoff, R. and Eppelt, S. and Zhou, Y. and Malik, P. and Seubert, M. and Bauer, T. and Rosenfeld, W. and Zhang, W. and Becher, C. and Weinfurter, H.},
  title   = {Entangling Single Atoms Over 33 km Telecom Fibre},
  journal = {Nature},
  volume  = {607},
  pages   = {69},
  year    = {2022},
  doi     = {10.1038/s41586-022-04764-4}
}

@article{Ritter2012,
  author  = {Ritter, S. and N{\"o}lleke, C. and Hahn, C. and Reiserer, A. and Neuzner, A. and Uphoff, M. and M{\"u}cke, M. and Figueroa, E. and Bochmann, J. and Rempe, G.},
  title   = {An Elementary Quantum Network of Single Atoms in Optical Cavities},
  journal = {Nature},
  volume  = {484},
  pages   = {195},
  year    = {2012},
  doi     = {10.1038/nature11023}
}

@article{Specht2011,
  author  = {Specht, H. P. and N{\"o}lleke, C. and Reiserer, A. and Uphoff, M. and Figueroa, E. and Ritter, S. and Rempe, G.},
  title   = {A Single-Atom Quantum Memory},
  journal = {Nature},
  volume  = {473},
  pages   = {190},
  year    = {2011},
  doi     = {10.1038/nature09997}
}

@article{Kalb2015,
  author  = {Kalb, N. and Reiserer, A. and Ritter, S. and Rempe, G.},
  title   = {Heralded Storage of a Photonic Quantum Bit in a Single Atom},
  journal = {Phys. Rev. Lett.},
  volume  = {114},
  pages   = {220501},
  year    = {2015},
  doi     = {10.1103/PhysRevLett.114.220501}
}

@article{Delteil2017,
  author  = {Delteil, A. and Sun, Z. and F{\"a}lt, S. and Imamo{\u{g}}lu, A.},
  title   = {Realization of a Cascaded Quantum System: Heralded Absorption of a Single Photon Qubit by a Single-Electron Charged Quantum Dot},
  journal = {Phys. Rev. Lett.},
  volume  = {118},
  pages   = {177401},
  year    = {2017},
  doi     = {10.1103/PhysRevLett.118.177401}
}

@article{Bhaskar2020,
  author  = {Bhaskar, M. K. and Riedinger, R. and Machielse, B. and Levonian, D. S. and Nguyen, C. T. and Knall, E. N. and Park, H. and Englund, D. and Lon{\v{c}}ar, M. and Sukachev, D. D. and Lukin, M. D.},
  title   = {Experimental Demonstration of Memory-Enhanced Quantum Communication},
  journal = {Nature},
  volume  = {580},
  pages   = {60},
  year    = {2020},
  doi     = {10.1038/s41586-020-2103-5}
}

@article{Jiang2009,
  author  = {Jiang, L. and Hodges, J. S. and Maze, J. R. and Maurer, P. and Taylor, J. M. and Cory, D. G. and Hemmer, P. R. and Walsworth, R. L. and Yacoby, A. and Zibrov, A. S. and Lukin, M. D.},
  title   = {Repetitive Readout of a Single Electronic Spin via Quantum Logic with Nuclear Spin Ancillae},
  journal = {Science},
  volume  = {326},
  pages   = {267},
  year    = {2009},
  doi     = {10.1126/science.1176496}
}

@article{Neumann2010,
  author  = {Neumann, P. and Beck, J. and Steiner, M. and Rempp, F. and Fedder, H. and Hemmer, P. R. and Wrachtrup, J. and Jelezko, F.},
  title   = {Single-Shot Readout of a Single Nuclear Spin},
  journal = {Science},
  volume  = {329},
  pages   = {542},
  year    = {2010},
  doi     = {10.1126/science.1189075}
}

@article{Reiserer2016,
  author  = {Reiserer, A. and Kalb, N. and Blok, M. S. and van Bemmelen, K. J. M. and Taminiau, T. H. and Hanson, R. and Twitchen, D. J. and Markham, M.},
  title   = {Robust Quantum-Network Memory Using Decoherence-Protected Subspaces of Nuclear Spins},
  journal = {Phys. Rev. X},
  volume  = {6},
  pages   = {021040},
  year    = {2016},
  doi     = {10.1103/PhysRevX.6.021040}
}

@article{Kalb2018,
  author  = {Kalb, N. and Humphreys, P. C. and Slim, J. J. and Hanson, R.},
  title   = {Dephasing Mechanisms of Diamond-Based Nuclear-Spin Memories for Quantum Networks},
  journal = {Phys. Rev. A},
  volume  = {97},
  pages   = {062330},
  year    = {2018},
  doi     = {10.1103/PhysRevA.97.062330}
}

@article{Maze2011,
  author  = {Maze, J. R. and Gali, A. and Togan, E. and Chu, Y. and Trifonov, A. and Kaxiras, E. and Lukin, M. D.},
  title   = {Properties of Nitrogen-Vacancy Centers in Diamond: The Group Theoretic Approach},
  journal = {New J. Phys.},
  volume  = {13},
  pages   = {025025},
  year    = {2011},
  doi     = {10.1088/1367-2630/13/2/025025}
}

@article{Doherty2011,
  author  = {Doherty, M. W. and Manson, N. B. and Delaney, P. and Hollenberg, L. C. L.},
  title   = {The Negatively Charged Nitrogen-Vacancy Centre in Diamond: The Electronic Solution},
  journal = {New J. Phys.},
  volume  = {13},
  pages   = {025019},
  year    = {2011},
  doi     = {10.1088/1367-2630/13/2/025019}
}

@article{Reyes2022,
  author  = {Reyes, R. and Nakazato, T. and Imaike, N. and Matsuda, K. and Tsurumoto, K. and Sekiguchi, Y. and Kosaka, H.},
  title   = {Complete Bell State Measurement of Diamond Nuclear Spins under a Complete Spatial Symmetry at Zero Magnetic Field},
  journal = {Appl. Phys. Lett.},
  volume  = {120},
  pages   = {194002},
  year    = {2022},
  doi     = {10.1063/5.0088155}
}

@article{Hadden2010,
  author  = {Hadden, J. P. and Harrison, J. P. and Stanley-Clarke, A. C. and Marseglia, L. and Ho, Y.-L. D. and Patton, B. R. and O'Brien, J. L. and Rarity, J. G.},
  title   = {Strongly Enhanced Photon Collection from Diamond Defect Centers under Microfabricated Integrated Solid Immersion Lenses},
  journal = {Appl. Phys. Lett.},
  volume  = {97},
  pages   = {241901},
  year    = {2010},
  doi     = {10.1063/1.3519847}
}

@article{Siyushev2010,
  author  = {Siyushev, P. and Kaiser, F. and Jacques, V. and Gerhardt, I. and Bischof, S. and Fedder, H. and Dodson, J. and Markham, M. and Twitchen, D. and Jelezko, F. and Wrachtrup, J.},
  title   = {Monolithic Diamond Optics for Single Photon Detection},
  journal = {Appl. Phys. Lett.},
  volume  = {97},
  pages   = {241902},
  year    = {2010},
  doi     = {10.1063/1.3519849}
}

@article{Robledo2011,
  author  = {Robledo, L. and Childress, L. and Bernien, H. and Hensen, B. and Alkemade, P. F. A. and Hanson, R.},
  title   = {High-Fidelity Projective Read-Out of a Solid-State Spin Quantum Register},
  journal = {Nature},
  volume  = {477},
  pages   = {574},
  year    = {2011},
  doi     = {10.1038/nature10401}
}

@article{Ruf2021,
  author  = {Ruf, M. and Weaver, M. J. and van Dam, S. B. and Hanson, R.},
  title   = {Resonant Excitation and Purcell Enhancement of Coherent Nitrogen-Vacancy Centers Coupled to a Fabry-Perot Microcavity},
  journal = {Phys. Rev. Appl.},
  volume  = {15},
  pages   = {024049},
  year    = {2021},
  doi     = {10.1103/PhysRevApplied.15.024049}
}

@article{Lindner2009,
  author  = {Lindner, N. H. and Rudolph, T.},
  title   = {Proposal for Pulsed On-Demand Sources of Photonic Cluster State Strings},
  journal = {Phys. Rev. Lett.},
  volume  = {103},
  pages   = {113602},
  year    = {2009},
  doi     = {10.1103/PhysRevLett.103.113602}
}

@article{Buterakos2017,
  author  = {Buterakos, D. and Barnes, E. and Economou, S. E.},
  title   = {Deterministic Generation of All-Photonic Quantum Repeaters from Solid-State Emitters},
  journal = {Phys. Rev. X},
  volume  = {7},
  pages   = {041023},
  year    = {2017},
  doi     = {10.1103/PhysRevX.7.041023}
}

@article{Russo2018,
  author  = {Russo, A. and Barnes, E. and Economou, S. E.},
  title   = {Photonic Graph State Generation from Quantum Dots and Color Centers for Quantum Communications},
  journal = {Phys. Rev. B},
  volume  = {98},
  pages   = {085303},
  year    = {2018},
  doi     = {10.1103/PhysRevB.98.085303}
}

@article{Bradley2022,
  author  = {Bradley, C. E. and de Bone, S. W. and M{\"o}ller, P. F. W. and Baier, S. and Degen, M. J. and Loenen, S. J. H. and Bartling, H. P. and Markham, M. and Twitchen, D. J. and Hanson, R. and Elkouss, D. and Taminiau, T. H.},
  title   = {Robust Quantum-Network Memory Based on Spin Qubits in Isotopically Engineered Diamond},
  journal = {npj Quantum Inf.},
  volume  = {8},
  pages   = {122},
  year    = {2022},
  doi     = {10.1038/s41534-022-00637-w}
}

\clearpage
\onecolumngrid

\begin{center}
    \textbf{\large Supplementary Information}
\end{center}

\renewcommand{\theequation}{S\arabic{equation}}
\setcounter{equation}{0}

\newcounter{suppnote}
\renewcommand{\thesuppnote}{\arabic{suppnote}}
\newcommand{\supplementarynote}[2]{%
    \refstepcounter{suppnote}%
    \subsection*{Supplementary Note \thesuppnote: #1}%
    \label{#2}%
}

\supplementarynote{Nuclear-spin coherence during repeated excitation}{supp:coherence}

In the orbital ground state, the nitrogen nuclear spin interacts with the
electron spin through the hyperfine interaction
\begin{equation}
H_{\mathrm{hf}} = \hbar\,\omega_{A}\, S_{z}\otimes I_{z},
\label{eq:sup1_hamiltonian}
\end{equation}
where $S_{z}$ and $I_{z}$ take the values $\pm 1$. The corresponding
time-evolution operator is
\begin{equation}
U(t) = \cos(\omega_{A}t)\; 1_{\mathrm{S}}\otimes 1_{\mathrm{N}}
     - i\sin(\omega_{A}t)\; S_{z}\otimes I_{z},
\label{eq:sup1_evolution}
\end{equation}
which becomes a non-entangling operation periodically at
\begin{equation}
t = \frac{n\pi}{2\omega_{A}},
\label{eq:sup1_condition}
\end{equation}
with integer $n$. The measured hyperfine frequency
$\omega_{A}/2\pi = 2.19~\mathrm{MHz}$ gives a period of $114~\mathrm{ns}$
between consecutive non-entangling times. A timing offset from this
condition causes phase relaxation of the nuclear spin once the electron
spin is traced out.

The hyperfine interaction in the orbital excited state $\ket{A_{2}}$ is
effectively suppressed by the spin--orbit interaction. Under this
approximation, the stochastic spread of the relaxation time from the
excited state to the ground state makes the effective duration of the
ground-state hyperfine interaction fluctuate. Additional phase relaxation
of the nuclear spin therefore occurs even when the mean timing is set to
the non-entangling condition.

We model the decay of the $\ket{+}_{\mathrm{N}}$ state fidelity under
repeated excitation using the timing offset $\Delta t$ and an additional
decay factor $\kappa$ per excitation,
\begin{equation}
F = \frac{1}{2}
    \left\{ 1 + \kappa^{N}\left[\cos(2\omega_{A}\Delta t)\right]^{N-1} \right\}.
\label{eq:sup1_fidelity}
\end{equation}
The factor $\kappa^{N}$ accumulates over the $N$ excitation events, and the
factor $\cos(2\omega_{A}\Delta t)$ accumulates over the $N-1$
free-precession intervals that separate them. Fitting Fig.~2b with $N=11$,
where $\Delta t$ is the offset of $t_{\mathrm{rep}}$ from the nearest
non-entangling condition, gives
\begin{equation}
\Delta t = 1.6~\mathrm{ns},
\qquad
\cos(2\omega_{A}\Delta t) = 0.999
\label{eq:sup1_offset}
\end{equation}
for $t_{\mathrm{rep}} = 3.2~\mathrm{\mu s}$. Fitting Fig.~2c with this
offset fixed gives
\begin{equation}
\kappa = 0.961 .
\label{eq:sup1_kappa}
\end{equation}

Separately, the relative photon collection efficiency shown in the same
panel is fitted to a saturating exponential
\begin{equation}
E(N) = A_{\mathrm{eff}}\left(1 - e^{-mN}\right).
\label{eq:sup1_efficiency}
\end{equation}
where $A_{\mathrm{eff}}$ and $m$ are fitting parameters.

\supplementarynote{Quantum process tomography}{supp:tomography}
To characterize the repeater operation, we reconstruct the output-photon polarization state conditioned on a successful Bell-state measurement between the electron and nuclear spins.
For an input polarization state $\ket{\nu}_\mathrm{P_{in}}$, with $\rho^\nu_\mathrm{in}=\ket{\nu}_\mathrm{P_{in}}\!\bra{\nu}$, let $\rho^\nu_{\mathrm{P_{out}},\mathrm{S},\mathrm{N}}$ denote the joint state of the output photon, electron spin, and nuclear spin at the time of emission.
We denote the projector onto the measured output polarization state $\ket{\mu}_\mathrm{P_{out}}$ by $\Pi_\mu^\mathrm{P_{out}}=\ket{\mu}_\mathrm{P_{out}}\!\bra{\mu}$, and the projector associated with the successful Bell-state outcome by $\Pi_\mathrm{Bell}^{\mathrm{S,N}}=\ket{\Phi^+}_\mathrm{S,N}\!\bra{\Phi^+}$.
Let $p_\mu$ be the event that the output photon is detected in polarization $\mu$, and let $B$ be the event that the Bell-state measurement succeeds.
The probability of detecting polarization $\mu$ conditioned on Bell-measurement success is then
\begin{align}
    P_\nu(p_\mu\mid B)
    &= \frac{\mathrm{Tr}\!\left[
        \left(\Pi_\mu^\mathrm{P_{out}}\otimes\Pi_\mathrm{Bell}^{\mathrm{S,N}}\right)
        \rho^\nu_{\mathrm{P_{out}},\mathrm{S},\mathrm{N}}
        \right]}
        {\mathrm{Tr}\!\left[
        \left(I_\mathrm{P_{out}}\otimes\Pi_\mathrm{Bell}^{\mathrm{S,N}}\right)
        \rho^\nu_{\mathrm{P_{out}},\mathrm{S},\mathrm{N}}
        \right]} \nonumber\\
    &= \frac{P_\nu(p_\mu\cap B)}{P_\nu(B)} .
\end{align}
Conversely, the Bell-measurement success probability conditioned on detection in polarization $\mu$ is
\begin{align}
    P_\nu(B\mid p_\mu)
    &= \frac{\mathrm{Tr}\!\left[
        \left(\Pi_\mu^\mathrm{P_{out}}\otimes\Pi_\mathrm{Bell}^{\mathrm{S,N}}\right)
        \rho^\nu_{\mathrm{P_{out}},\mathrm{S},\mathrm{N}}
        \right]}
        {\mathrm{Tr}\!\left[
        \left(\Pi_\mu^\mathrm{P_{out}}\otimes I_\mathrm{S,N}\right)
        \rho^\nu_{\mathrm{P_{out}},\mathrm{S},\mathrm{N}}
        \right]} \nonumber\\
    &= \frac{P_\nu(p_\mu\cap B)}{P_\nu(p_\mu)} .
\end{align}
Thus, Bayes' rule gives
\begin{equation}
    P_\nu(p_\mu\mid B)
    = \frac{P_\nu(B\mid p_\mu)P_\nu(p_\mu)}{P_\nu(B)} .
\end{equation}

The two factors in the numerator are obtained independently from the experimental counts.
For input $\nu$ and measurement (polarizer) setting $\mu$, they are estimated as
\begin{equation}
    P_\nu(p_\mu)
    = \frac{N_{\nu\mu}^\mathrm{det}}
    {N_{\nu}^\mathrm{abs}\,\eta_\mathrm{det}},
    \qquad
    P_\nu(B\mid p_\mu)
    = \frac{N_{\nu\mu}^\mathrm{raw}}{N_{\nu\mu}^\mathrm{ref}} .
\end{equation}
Here, $N_{\nu\mu}^\mathrm{abs}$ is the number of absorption-herald events, and $N_{\nu\mu}^\mathrm{det}$ is the number of detected output photons.
The factor $\eta_\mathrm{det}$ is the total detection efficiency per absorption herald, including excitation, emission, optical coupling and transmission, and detector efficiencies that are independent of the measured polarization.
It is distinct from the single-emission extraction efficiency $\eta_\mathrm{emis}$ defined in Supplementary Note~\ref{supp:efficiency}.
$N_{\nu\mu}^\mathrm{raw}$ is the total photon count accumulated during the experiment, obtained by mapping the population associated with the target Bell state onto the electron state $\ket{0}_\mathrm{S}$ and optically reading out that population, and $N_{\nu\mu}^\mathrm{ref}$ is the corresponding reference count for a fully initialized electron state $\ket{0}_\mathrm{S}$.

The denominator $P_\nu(B)$ need not be measured separately.
For two orthogonal measured outcomes $\mu$ and $\mu'$ forming a complete polarization basis, $P_\nu(p_\mu\mid B)+P_\nu(p_{\mu'}\mid B)=1$.
Consequently, the desired conditional probability is obtained by normalizing the two joint-event weights:
\begin{align}
    P_\nu(p_\mu\mid B)
    &= \frac{P_\nu(B\mid p_\mu)P_\nu(p_\mu)}
    {P_\nu(B\mid p_\mu)P_\nu(p_\mu)
    +P_\nu(B\mid p_{\mu'})P_\nu(p_{\mu'})} \nonumber\\
    &= \frac{w_{\nu\mu}}{w_{\nu\mu}+w_{\nu\mu'}},
\end{align}
where
\begin{equation}
    w_{\nu\mu}
    = \frac{N_{\nu\mu}^\mathrm{raw}}{N_{\nu\mu}^\mathrm{ref}}
      \frac{N_{\nu\mu}^\mathrm{det}}{N_{\nu\mu}^\mathrm{abs}} .
\end{equation}
The polarization-independent efficiency $\eta_\mathrm{det}$ cancels in this normalization, so the conditional output probabilities follow directly from the measured counts.

To perform the state and process tomography, we now restrict the input state $\nu$ to the six Pauli eigenstates, $\ket{\pm}$, $\ket{\pm i}$, and $\ket{\pm 1}$.
Accordingly, we choose the measurement setting $\mu$ to project onto these basis states.
For a given Pauli operator $\sigma_m$ ($m=1,2,3$), let $\mu$ and $\mu'$ denote its $+1$ and $-1$ eigenstates.
The expectation value is then evaluated from the conjugate pair as
\begin{equation}
    \langle\sigma_m\rangle_\nu
    = P_\nu(p_\mu\mid B)-P_\nu(p_{\mu'}\mid B).
\end{equation}
The emitted-photon density matrix follows by linear inversion,
\begin{equation}
    \rho^\nu_\mathrm{out}
    = \frac{1}{2}\left(\sigma_0+\sum_{m=1}^{3}\langle\sigma_m\rangle_\nu\,\sigma_m\right),
\end{equation}
with $\sigma_0=I$.
The repeater channel is represented by a process matrix $\chi$ in the same Pauli basis,
\begin{equation}
    \mathcal{E}_\chi(\rho^\nu_\mathrm{in})
    = \sum_{m,n=0}^{3}\chi_{mn}\,\sigma_m\rho^\nu_\mathrm{in}\sigma_n ,
\end{equation}
and is determined by minimizing
\begin{equation}
    \sum_\nu\left\lVert\rho^\nu_\mathrm{out}-\mathcal{E}_\chi(\rho^\nu_\mathrm{in})\right\rVert_\mathrm{F}^{2}
\end{equation}
subject to
\begin{equation}
    \chi\succeq0,\qquad \sum_{m,n=0}^{3}\chi_{mn}\,\sigma_n\sigma_m=\sigma_0 .
\end{equation}
These constraints ensure that the reconstructed process is a physically valid quantum channel.

\supplementarynote{Effect of crystal strain on the excited state}{supp:strain}
The perturbation Hamiltonian of the NV-center excited state induced by crystal strain is given by\cite{Maze2011,Doherty2011}
\begin{equation}
\begin{split}
    H_{strain} = \delta_x\big(&\ket{E_x}\bra{E_x} - \ket{E_y}\bra{E_y} + \ket{A_1}\bra{E_1} \\
    &+ \ket{E_1}\bra{A_1} + \ket{A_2}\bra{E_2} + \ket{E_2}\bra{A_2}\big).
\end{split}
\end{equation}
The effective magnitude of the strain experienced by the NV center can be estimated from the frequency splitting between the $E_x$ and $E_y$ optical resonances, which is given as \qty{1.2}{\giga\hertz} by the PLE spectrum on this NV center.
From this estimate, the state fidelity of the perturbed $\ket{A_2}_\mathrm{S,L}$ level relative to the unperturbed eigenstate is evaluated to be \qty{99.5}{\percent}.
Therefore, crystal strain is far too small to account for the observed fidelity degradation in the present experiment, which is instead dominated by the excited-state relaxation jitter described in Supplementary Note~\ref{supp:coherence}.

\supplementarynote{Efficiency advantage of an absorption--emission repeater}{supp:efficiency}
We estimate the efficiency advantage of a chain of identical absorption--emission repeater nodes over direct transmission.
Since every node performs the same operation, it is sufficient to evaluate a single segment of the chain, in which a quantum state already stored in the memory of node $k$ is transferred to the memory of node $k+1$.
Let $\eta_\mathrm{emis}$ denote the emission efficiency of a node, $\eta_\mathrm{link}$ the transmission efficiency of the optical link between neighboring nodes, and $\eta_\mathrm{abs}$ the absorption efficiency of a node.
Throughout this estimate we assume that the nuclear-spin quantum memory suffers no decoherence during the RUS attempts, and that the absorption and emission phases of neighboring nodes are scheduled such that no node performs both operations at the same time.

A single RUS attempt consists of an optical excitation at node $k$, propagation of the emitted ZPL photon to node $k+1$, and a heralded absorption event at node $k+1$, whose outcome is communicated back to node $k$ before the next attempt is made.
The success probability of a single attempt is
\begin{equation}
    p_\mathrm{seg} = \eta_\mathrm{emis}\,\eta_\mathrm{link}\,\eta_\mathrm{abs} .
\end{equation}

Because the memory retains the stored state between attempts, the emission is retried up to $N_\mathrm{max}$ times for a single stored state, and the state is transferred to the neighboring node with probability
\begin{equation}
    P_\mathrm{seg} = 1 - \left(1-p_\mathrm{seg}\right)^{N_\mathrm{max}} .
\end{equation}
Inserting a node into an optical link removes the transmission loss of the bypassed section and introduces instead the internal loss of the node.
An approximate criterion for an advantage over direct transmission is therefore
\begin{equation}
    P_\mathrm{seg} > \eta_\mathrm{link} .
\end{equation}
Once this condition is satisfied, every additional node contributes the same constant factor $P_\mathrm{seg}/\eta_\mathrm{link}$ to the overall efficiency, and the advantage over direct transmission accumulates exponentially with the number of nodes.

In practice, $N_\mathrm{max}$ is bounded by the coherence of the nuclear-spin memory, and the emission step is terminated once the maximum number of attempts has been reached.
The RUS protocol retains its advantage as long as $N_\mathrm{max}$ is large enough for $P_\mathrm{seg}$ to approach unity while the memory fidelity remains high.
In the present experiment $N_\mathrm{max}=10$ preserves the nuclear-spin state with an average fidelity of \qty{87}{\percent}, which by itself does not yet outperform direct transmission.
Since the end-to-end fidelity degrades further with increasing number of segments, a more stable quantum memory or entanglement purification will be an important functionality for future implementations.

The present estimate concerns the transfer efficiency rather than the generation rate.
Direct transmission of a single photon can be attempted at the full repetition rate of the photon source, whereas each RUS attempt must await the heralding signal returning from the neighboring node at a distance $L$, so that the attempt interval cannot be shorter than the round-trip time $2L/c$.
The two schemes are therefore not compared here on an equal basis in terms of rate.
The same round-trip time is nevertheless incurred in direct transmission whenever heralding is required.
The efficiency also remains the relevant figure of merit when a quantum state must be delivered without loss, for example when transferring a state that cannot be regenerated on demand.

\end{document}